\documentclass[aps,prb,twocolumn,amsmath,superscriptaddress]{revtex4-2}

\newcommand{\bea}{\begin{eqnarray}}
\newcommand{\eea}{\end{eqnarray}}
\newcommand{\beq}{\begin{equation}}
\newcommand{\eeq}{\end{equation}}

\newcommand{\dd}{d^{\dagger}}
\newcommand{\dnd}{d^{\phantom{\dagger}}}

\usepackage[urlcolor=blue,colorlinks=true,citecolor=blue,linkcolor=blue,pdfstartview={FitH},bookmarks=false]{hyperref}
\usepackage{graphicx}
\usepackage{longtable}
\usepackage{epsfig}
\usepackage{dcolumn}
\usepackage{bm}
\usepackage{amssymb}
\usepackage{multirow}
\usepackage{times,xcolor}
\usepackage{hyperref}
\usepackage{braket}
\usepackage{comment}

\begin{document}

\title{Hubbard subbands and superconductivity in the infinite-layer nickelate}

\author{Tharathep Plienbumrung}
\author{Maria Daghofer}
\affiliation{\mbox{Institute for Functional Matter and Quantum Technologies,
University of Stuttgart, Pfaffenwaldring 57, D-70550 Stuttgart, Germany}}
\affiliation{\mbox{Center for Integrated Quantum Science and Technology,
University of Stuttgart, Pfaffenwaldring 57, D-70550 Stuttgart, Germany}}

\author{Michael T. Schmid}
\affiliation{\mbox{Waseda Research Institute for Science and Engineering,
Waseda University, Okubo, Shinjuku, Tokyo, 169-8555, Japan}}

\author{Andrzej M. Ole\'s$\,$}
\email{Corresponding author: a.m.oles@fkf.mpi.de}
\affiliation{\mbox{Max Planck Institute for Solid State Research,
             Heisenbergstrasse 1, D-70569 Stuttgart, Germany} }
\affiliation{\mbox{Institute of Theoretical Physics, Jagiellonian University,
             Profesora Stanis\l{}awa \L{}ojasiewicza 11, PL-30348 Krak\'ow, Poland}}

\begin{abstract}
An effective two-dimensional two-band model for infinite-layer
nickelates consists of bands obtained from $d_{x^2-y^2}$ and $s$--like
orbitals. We investigate whether it could be mapped onto a single-band
Hubbard model and the filling of Hubbard bands. We find that both
one-band physics and a Kondo-lattice regime emerge from the same
two-orbital model, depending on the strength of electronic correlations
and the filling of the itinerant $s$-band. Next we investigate
one-particle excitations by changing the screening.
First, for weak screening the strong correlations push electrons out
of the $s$-band so that the undoped nickelate is similar to a cuprate.
Second, for strong screening the $s$ and $d_{x^2-y^2}$ bands are both
partly filled and weakly coupled. Particularly in this latter regime
mapping to a one-band model gives significant spectral weight transfer
between the Hubbard subbands.
Finally we show how the symmetry of superconducting phases depends on 
the interaction parameters and determine the regions of $d$-wave or 
$s$-wave symmetry.
\end{abstract}

\date{\today}

\maketitle


\section{Introduction}


A few years ago, superconductivity was reported in
infinite-layer NdNiO$_2$ thin films with Sr doping \cite{Li19}. The
lattice structure shares similarities with cuprate superconductors,
with NiO$_2$ planes taking the place of  CuO$_2$ planes. While both
can be expected to be quite correlated and both show antiferromagnetic
(AFM) superexchange~\cite{Jia20,Tha21,Hep20,Lu21,Lin21}, there  are 
some microscopic differences. One is the lack of apical oxygens in the
Ni case, which affects crystal-field energies, the other is the
presence of dispersive rare-earth states close to the Fermi
level. Whether one starts from isolated NiO$_2$ layers
\cite{Jia20,Tha21,Hep20,Lu21,Lin21,Hu19,Zha20} or from
band-structure calculations
\cite{Jia19,Si20,Wu20,Kle21,Zha21,Bee21,Hig21,Bot22}, one  
expects that more than one orbital or band might be relevant.

\begin{figure}[t!]
	\includegraphics[width=\linewidth]{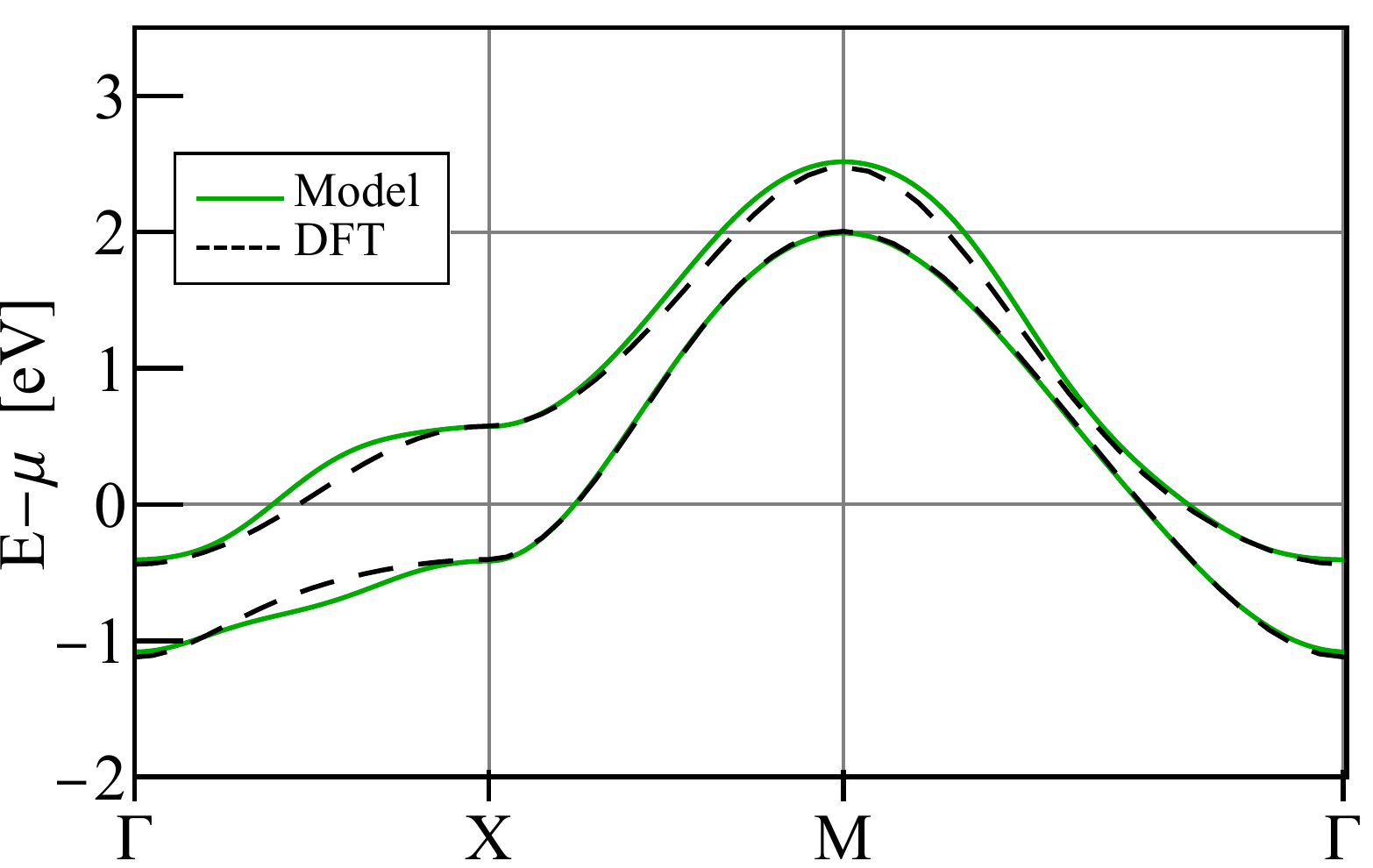}
\caption{\textcolor{black}{Non-interacting band structure of NdNiO$_2$:
DFT bands crossing the Fermi surface (black
dashed lines) and the 2D tight-binding model obtained by projecting a
Wannier fit onto the plane along 2D path (green solid lines).
The Fermi energy $E=\mu$ corresponds to the DFT electronic structure.
}}
	\label{fig:bands}
\end{figure}

While single-band~\cite{Kit20} and three-band~\cite{Wu20,Kre22} models 
have also been proposed, two bands cross the Fermi level, see Fig. 
\ref{fig:bands}, and many groups have accordingly investigated two-band
models~\cite{Nom19,Adh20,Gu20,LecX,Xie21,Kre22}. One of the bands
has a large contribution from the $x^2-y^2$ orbital at Ni and its
dispersion is nearly perfectly two-dimensional (2D). This band can be
expected to share features with the Cu-dominated band of the cuprates
to be rather correlated. In the second band, rare-earth states
hybridize with Ni apical
states, thus obtaining some Ni($d_{3z^2-r^2}$) and Ni($d_{xy}$)
character, however, its wave function has $s$-symmetry~\cite{Adh20},
and we denote it accordingly. Previous studies of various two-band
models have yielded a large variety of potential pairing symmetries
\cite{Hu19,Ole19}, among them $s$, $d$, and $s_\pm$-wave states
\cite{Wu20,Zha20,Kre22}, while a one-band scenario favors $d$-wave
\cite{Kit20}.

The $s$-like band lies mostly above the Fermi level, however, it forms
electron pockets around the $\Gamma$ and $A$ points in the Brillouin
zone. In the DFT band structure, the pockets account for $\approx7\,\%$
of the occupied states~\cite{Bot20}. Even without Sr-doping, these
electrons are thus missing from the $x^2-y^2$ band. When translating to
a cuprate scenario, it should be noted that $5\,\%$
of Sr doping suffices to destroy antiferromagnetism in
La$_2$CuO$_{4-y}$ \cite{Bud88}. We thus have to expect partly filled
bands, and at least one of them is correlated.

The purpose of this paper is to investigate the evolution of the bands
shown in Fig.~\ref{fig:bands} with crystal-field splittings and electron
correlations in both bands. Thereby we investigate to what extent
multi-band effects come into play in nickelates. It is important to
realize that partial filling of the strongly correlated $x^2-y^2$
orbitals means that the electronic spectral weight may be transferred
from the upper Hubbard band (UHB) to the lower Hubbard band (LHB) above
the Fermi energy. The mechanism of such a weight transfer is well known
for the doped Hubbard model \cite{Esk91,Mei93}. Another mechanism which
promotes such a weight transfer is interaction screening that generates
finite filling within the weakly correlated orbitals of $s$-wave
symmetry. It is remarkable that both for very weak and for strong
screening the bands mostly decouple and the effective physics becomes
similar to a single Hubbard band~\cite{Kit20}. We then find a Mott
insulator (doped band with potential $d$-wave pairing) for strong 
(weak) correlations. Correlation strength thus emerges as an
important factor in the description of nickelate superconductors.

The remaining of this paper is organized as follows. The two-band model
arises from the electronic structure calculations as described in Sec.
\ref{sec:two}. Electronic interactions are given by two Kanamori
parameters for an atom $\alpha$: $\{U_{\alpha},J_H\}$, and we discuss 
their screening in Sec. \ref{sec:int}. Below we
investigate a very wide range of potential regimes going from weak to
very strong correlations in Sec. \ref{sec:numa}. There we show how the
density of states changes with screening for strong and weak
interactions in Sec. \ref{sec:numb}. Finally, we present the essential
AFM and superconducting (SC) phases for various regimes of screening in
Sec. \ref{sec:sc}. The paper is concluded in Sec. \ref{sec:summa}.

\section{Two-band model and Methods}
\label{sec:mod}
\subsection{Kinetic energy}
\label{sec:two}

We start from the kinetic energy in the electronic structure. The DFT
band structure, see Fig. \ref{fig:bands}, is calculated with
{\sc Quantum} {\sc Espresso} code \cite{Gia09,Gia17,Gia20} using a
plane-wave pseudopotential method \cite{Dal14}. As discussed in Ref.
\cite{Max}, many models can be constructed that differ in the 
shape of the apical $s$-like orbital. Since their hopping integrals,
given in~\cite{Max}, are nevertheless very similar, the kinetic energy
is not affected by this ambiguity in any physically relevant way.

The Wannier90 interface~\cite{Piz20} gives the parametrization,
\begin{align}
\label{eq1}
H_{\rm kin}&=\sum_{i,\lambda=d,s;\sigma}\epsilon_{\lambda}
\dd_{i\lambda\sigma}\dnd_{i\lambda\sigma}
+\sum_{ij,\{\lambda\mu\},\sigma}{t_{ij}^{\lambda\mu}
d^{\dagger}_{i\lambda\sigma}d^{}_{j\mu\sigma}},
\end{align}
where $\dnd_{i\lambda\sigma}$ ($\dd_{i\lambda\sigma}$) annihilates 
(creates) an electron  at site $i$ in orbital $\lambda=d,s$, with spin 
$\sigma$. ($d$ and $s$ refer to the two bands discussed above and shown
in Fig. 1.) Hopping parameters $t^{\lambda\mu}_{ij}$ and on-site 
energies $\epsilon_{\lambda}$ are given in Ref.~\cite{Max}.
We project these three-dimensional bands onto the $(x,y)$-plane, as
we are here mostly interested in the correlated $x^2-y^2$ states,
whose band is already quite 2D to start with~\cite{Max}.

\subsection{Interactions and screening effect}
\label{sec:int}

While the rather extended wave-function of (especially) the $s$-like
state might lead to longer-ranged Coulomb interactions, on-site terms
can be expected to dominate and we use (intraorbital and interorbital)
Coulomb elements of the form \cite{Tha21},
\begin{align}
\label{eq2}
\!H_{\rm int}&=
\sum_{i,\lambda=d,s}U_{\lambda}n_{i\lambda\uparrow}n_{i\lambda\downarrow}
 + \left(U'-\frac{J_{H}}{2}\right)\sum_{i}{n_{id}n_{is}} \\
& - 2J_{H}\sum_{i}\vec{S}_{id}\!\cdot\!\vec{S}_{is}
+\!J_H\sum_i\!\left(\dd_{id\uparrow}\dd_{id\downarrow}
\dnd_{is\downarrow}\dnd_{is\uparrow}+{\rm H.c.}\right)\!.   \nonumber
\end{align}
$n_{i\lambda\sigma}^{}$ is the electron number operator at site $i$,
in orbital $\lambda$ and for spin $\sigma$ and
$\{\vec{S}_{i\lambda}\}$ the  corresponding spin operator.
Intraorbital Coulomb repulsion $U_{\lambda}$ depends on the band index
$\lambda=d,s$. Hund's exchange is given by $J_H$
and interorbital repulsion $U'$ couples the bands.

Upper limits for the 'bare' $U_d$ and $J_0$ are given by their atomic
values $U_d\approx 8$ eV and $J_0\approx 1.2$ eV, as one might use in
modelling an insulating NiO$_2$ layer \cite{Jia20,Tha21}. However,
when projecting out oxygen states and using Wannier functions instead,
effective values have to be significantly reduced. In the case of $U_s$, 
atomic values for Ni cannot be even taken as a starting point, as the
$s$-orbital is mostly made up of rare-earth states and is not centered 
on a Ni site~\cite{Adh20}. The strong Nd($5d$) character and very
itinerant character of the $s$-bands suggests that their effective
interaction should be strongly screened, in fact more that can be 
expected for the $d$ states. One thus expects $U_d>U_s$, which we take 
into account in a phenomenological way via a screening parameter 
$\alpha\in [0,1]$, so that
\begin{align}\label{eq:U_J_alpha}
	U_s = \alpha U_d,\quad	J_H = \alpha J_0,\quad 	U' = U_s - 2J_H.
\end{align}
This parametrization provides the simplest approach to discuss the
interplay of a more and a less correlated band.

We then use Lanczos exact diagonalization to treat the full Hamiltonian
$H=H_{\rm kin}+H_{\rm int}$ on an eight-site square cluster standing 
for a NiO$_2$ plane. Orbital densities are analyzed following Ref. 
\cite{Max} for the two orbitals, $d$ and $s$. Below we summarize the 
evolution of the density of states in the correlated $d$ band and the 
accompanying $s$ band. We particularly focus on the circumstances 
favoring the electron transfer between the Hubbard subbands in the 
correlated band.

\section{Numerical results}

\subsection{Hubbard subbands}
\label{sec:numa}

\begin{figure}[t!]
	\includegraphics[width=\columnwidth]{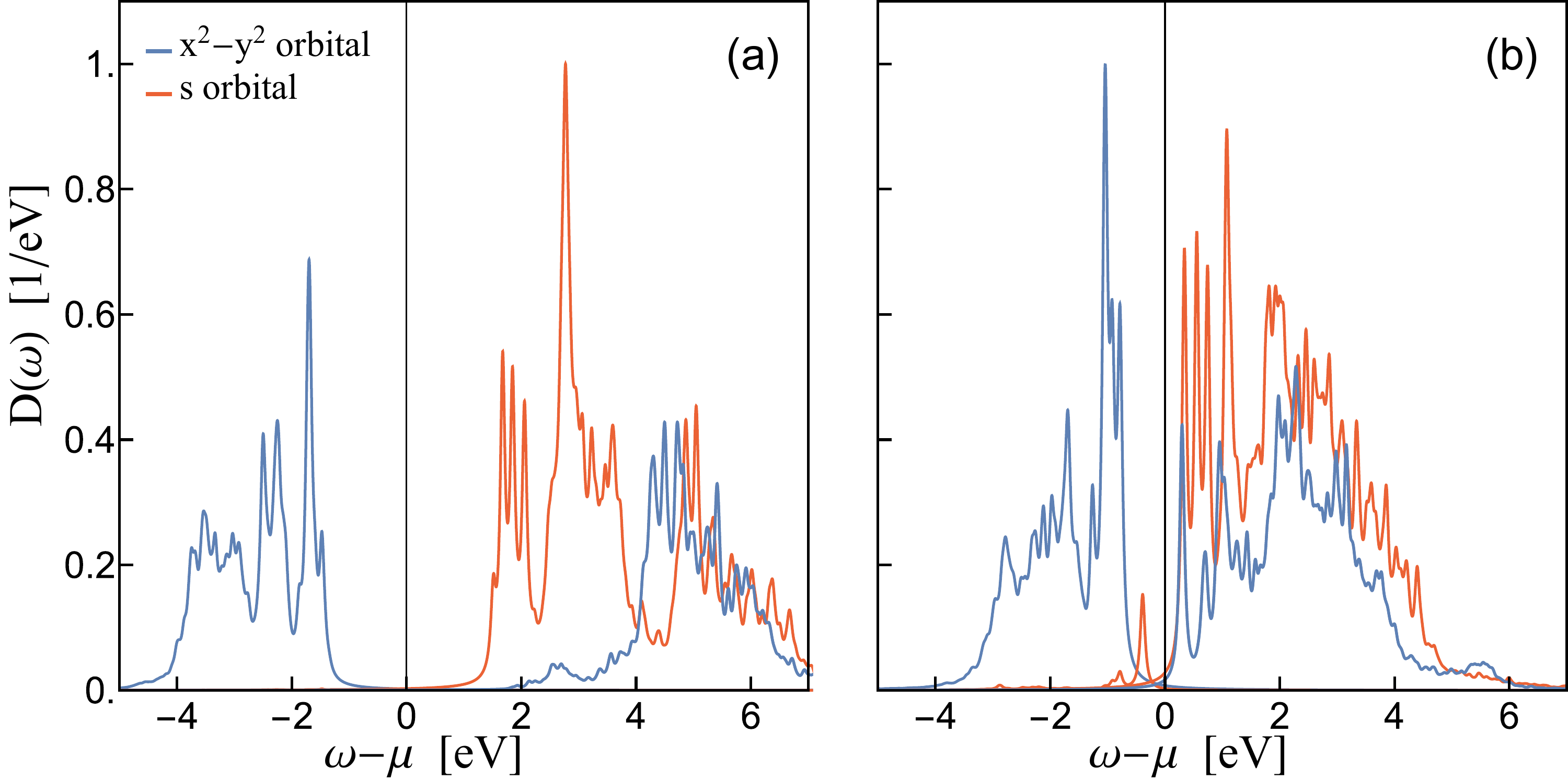}
	\caption{
Density of states $D(\omega)$ of undoped nickelate with unscreened
interactions ($\alpha=1$). Fermi energy is set to zero;
the orbital densities are normalized to one (per spin).
Intraorbital Coulomb interaction in Eq. \eqref{eq2} is selected at:
(a) $U_d=8$ eV and
\mbox{(b) $U_d=4$ eV.}}
	\label{fig:mu}
\end{figure}

To understand the occurrence of possible SC phase in infinite-layer
nickelates we consider first the one-particle spectra in the normal
phase. Figure \ref{fig:mu} shows the orbital-resolved density of states,
taking two values of the Coulomb interaction $U_d=8.0$ and $4.0$ eV.
The larger value is the same as interactions in cuprates \cite{Gra92}
and could be considered to be the upper limit; the lower value stands
for effective Coulomb interactions in the metallic state in nickelates
where Coulomb interactions are weaker. Here we begin with unscreened
interactions in the $s$ band, i.e., we take $\alpha=1.0$.

In both cases of large $U_d=8.0$ eV and moderate $U_d=4.0$ eV, one
finds a Mott insulator with two subbands separated by a gap, the
occupied LHB and the empty UHB. For large \mbox{$U_d=8.0$} eV, the gap 
is $\eqsim 3.5$ eV and the Fermi energy falls within the gap, see Fig. 
\ref{fig:mu}(a). This may be considered a textbook example of a Mott 
insulator. Then one finds also an AFM order in the LHB.

When $U_d=4$ eV, the gap in the correlated band decreases to less than
$\eqsim 1.0$ eV and the tail of the $s$ band falls below the Fermi
energy which still separates the occupied and unoccupied states of the
LHB, see Fig. \ref{fig:mu}(b). However, we should keep in mind that
the calculations are done for a finite system and we cannot exclude a
metallic phase in the thermodynamic limit. In any case, one finds a
small fraction of electrons occupying the $s$ states and these states
are just below the Fermi energy, see Fig. \ref{fig:mu}(b). Here the
correlated LHB band is less than half-filled and develops dynamics.
As a result, electron transfer from the UHB to the unoccupied part of
the LHB increases, and the total occupancy of the LHB exceeds 0.5. 
We conclude that the presence of the second more itinerant band is 
responsible for the electron transfer between the Hubbard subbands.

The next question to ask is where doped holes go in the quarter-filled
system. Before we have shown \cite{Tha21a} that three regimes emerge
for increasing screening as discussed below. The reduction of the
Coulomb interaction to $U_d=4.0$ eV is sufficient to cause the loss of
long-range AFM order in the $x^2-y^2$ orbital due to reduced electron
filling.

\subsection{One-particle spectral density}
\label{sec:numb}

First, in the weakly screened Mott insulator and for large $U_d$, the
ground state is AFM and holes naturally enter only the $x^2-y^2$ 
orbital. In contrast, the second regime is found at intermediate 
screening [$\alpha\simeq 0.5$], or for interactions that are reduced 
from the outset, see Fig.~\ref{fig:mu}(b). Finally, in the third regime 
of strong screening ($\alpha\sim 0.2$), hole doping occurs again into 
the $x^2-y^2$ orbital, with the $s$ electrons remaining unaffected
\cite{Tha21a}. This behaviour is presented in more detail below.

\begin{figure}[t!]
	\includegraphics[width=\columnwidth]{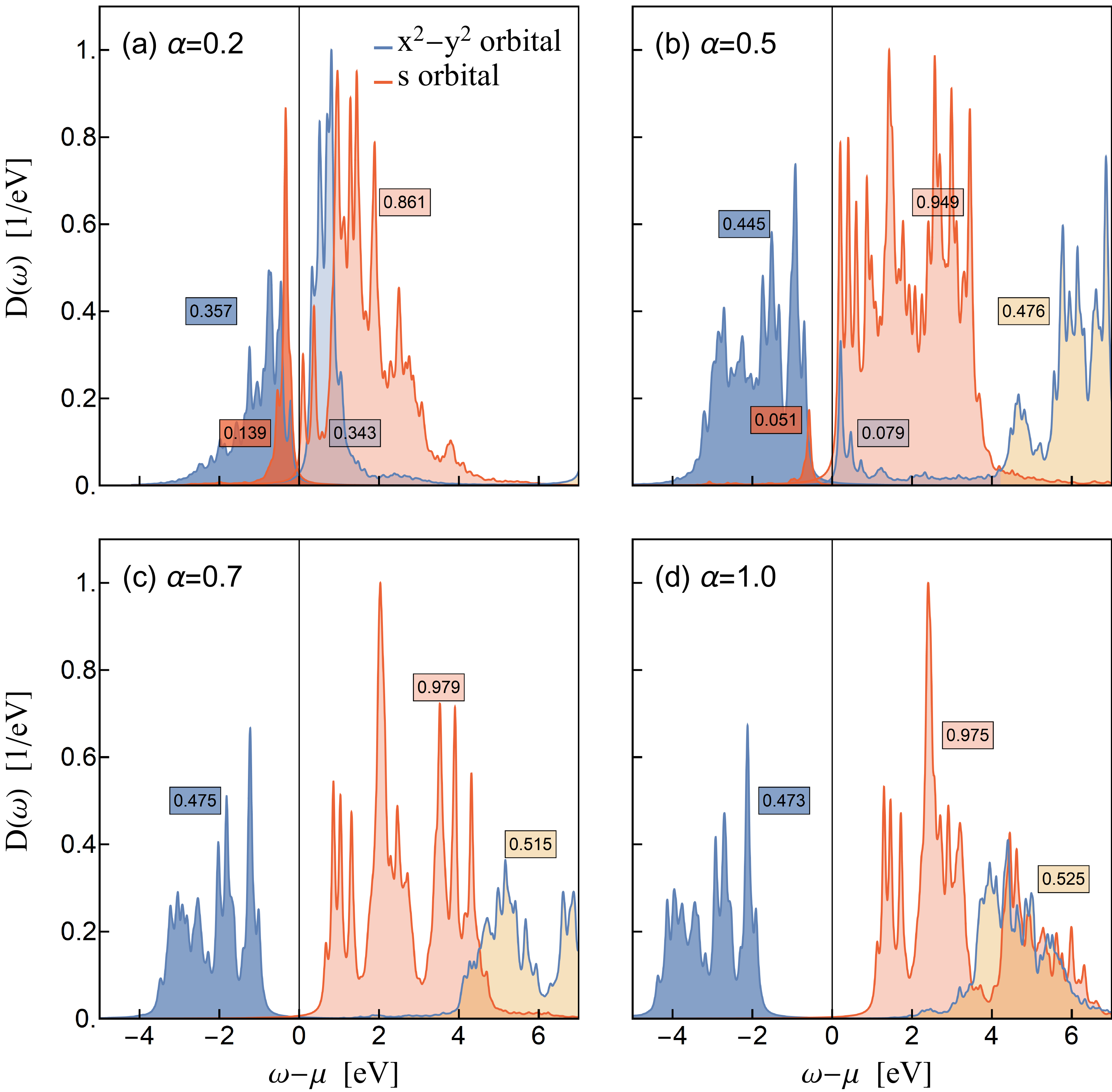}
	\caption{
Density of states $D(\omega)$ of undoped nickelate for $U_d=8.0$ eV
and for different values of screening $\alpha$.
}
	\label{fig:dos8}
\end{figure}

\begin{figure}[b!]
	\includegraphics[width=\columnwidth]{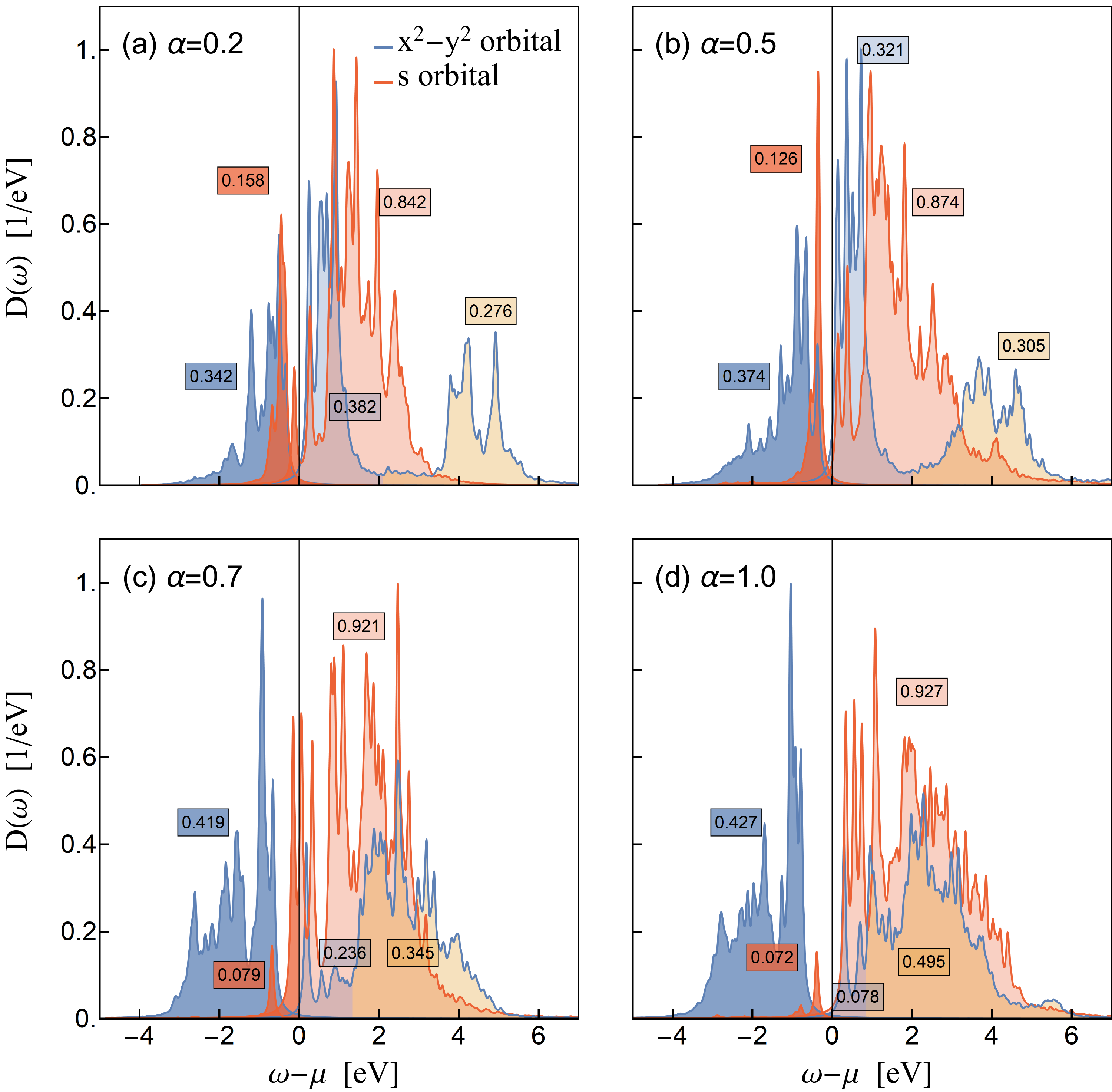}
	\caption{
Density of states $D(\omega)$ of undoped nickelate for $U_d=4.0$ eV
and for different values of screening $\alpha$.
}
	\label{fig:dos4}
\end{figure}

The different behaviour in the three regimes mentioned in Sec.
\ref{sec:numa} is also reflected in the single-particle spectra shown
in Fig.~\ref{fig:dos8}. Filling corresponds to doping with two holes
and twisted boundary conditions (TBC) are used to resolve more momenta \cite{Shi97,Poi91,Yin09}. Both for very strong $U_d=8.0$ eV, see Fig. 
\ref{fig:dos8}(a) and for moderate $U_d=4.0$ eV, see Fig. 
\ref{fig:dos4}(a), the correlations induce a gap in the 
$x^2-y^2$ band \cite{Lec20}. The lowest (occupied) states for electrons 
are in the $s$ band at reduced $U_d=4$ eV, i.e., both bands are partly 
filled. This can be seen in Fig.~\ref{fig:mu}(b), where we show the 
density of states for eight electrons (i.e., at quarter filling). Data 
were obtained by means of TBC, integrating over five sets of boundary 
conditions.

At strong screening when $\alpha=0.2$, the occupied states in the
$x^2-y^2$ band are rather similar for $U_d=8.0$ eV and $U_d=4.0$ eV,
except that the curvature of the occupied states changes along the
$(\pi,0)-(0,\pi)$ line. Also the values of $n_s$ and of the weight
transferred to the LHB are similar. Since this implies that
interactions $U'$ and $J_H$ between $d$ and $s$ states do not play here
a significant role, it supports the notion of a correlated (and doped)
$d$ band that is only affected by a metallic $s$ band via self-doping.

In contrast, for stronger correlations, i.e., weaker screening
$\alpha\in[0.5,0.7]$, the spectra shown in Figs.~\ref{fig:dos8}(b) and
\ref{fig:dos8}(c) are affected by $U_d$. All electrons are here in
the correlated $x^2-y^2$ states. It is remarkable that the occupied
states fall almost at the same energies, independently of whether
$U_d=8.0$ eV or $U_d=4.0$ eV [cf. Figs. \ref{fig:dos8} and
\ref{fig:dos4}]. However, splitting between $d$ and $s$ states is
clearly affected by $U_d$ (via $U'$ and $J_H$), which indicates that the
$s$- and $d$-bands are in this regime directly coupled, not only via
self-doping.

Analogous conclusions can be drawn from the undoped density of states
shown in Figs.~\ref{fig:dos8}(a) and \ref{fig:dos4}(a) are extremely
similar: In the regime of strong screening, both bands are partly
filled and results hardly depend on $U_d$ at all. In the intermediate
regime on the other hand, both bands are likewise partially filled. A
comparison between Figs.~\ref{fig:dos8} and \ref{fig:dos4} indicates
that the $s$ states being close to the Fermi level could be doped away. 
In this regime, results depend on $U_d$, indicating that correlations 
are here more important to describe low-energy features close to the 
Fermi level.

\begin{table}[t!]
\caption{
Electron densities $n_d$ and $n_s$ per spin obtained in the undoped
nickelate for screened interactions $(\alpha<1)$. The weight of the
LHB $w_{\rm LHB}$ in increased by the kinetic weight transfer from
the UHB \cite{EskPRL,Esk94}.
\label{tab:wei}
}
\begin{ruledtabular}
\begin{tabular}{ c c c c c c }
$U_d$ (eV) & $\alpha$ & $n_d$ & $n_s$ & $w_{\rm LHB}^{\;>}$ & $w_{\rm LHB}$ \\
\colrule
8.0 & 0.20 & 0.358 & 0.139 & 0.343 & 0.700   \\
    & 0.50 & 0.445 & 0.051 & 0.079 & 0.524   \\
    & 0.70 & 0.475 & 0.021 & 0.011 & 0.485   \\
    & 1.00 & 0.473 & 0.025 & 0.002 & 0.475   \\
\colrule
4.0 & 0.20 & 0.342 & 0.158 & 0.382 & 0.724   \\
    & 0.50 & 0.374 & 0.126 & 0.321 & 0.695   \\
    & 0.70 & 0.419 & 0.079 & 0.236 & 0.656   \\
    & 1.00 & 0.427 & 0.072 & 0.078 & 0.505   \\
\end{tabular}
\end{ruledtabular}
\end{table}

Interestingly, the screening increases the density of $s$ electrons and
simultaneously the density in the correlated band $n_d$ decreases as in 
the undoped case the constraint $n_d+n_s=1$ is satisfied. This
makes the LHB less than half-filled and considerable spectral weight is
transferred from the UHB to the unoccupied part of the LHB (i.e., above
the Fermi energy~$\mu$). The mechanism of such a spectral weight 
transfer is well known in the partly filled Hubbard model 
\cite{Esk91,Mei93} and explains why the weight of the LHB exceeds
eventually 0.5 per spin. Here doping in the Mott insulator is mimicked 
by the partial filling of the $s$ band. The largest transfer of spectral 
weight is found at $U_d=4.0$ eV and $\alpha=0.2$, see Table I. The UHB 
forms only in the correlated $x^2-y^2$ band and Hubbard subbands are
absent within the $s$ band even at $U_d=8.0$ eV.

The regimes of weak and strong screening differ qualitatively. The
number of correlated electrons $n_d$ is close to $n_d=0.5$ for weak
screening but decreases rapidly for large screening. As a result, the
total weight in the LHB $w_{\rm LHB}$ increases somewhat above $0.7$
both for $U_d=8.0$ and $U_d=4.0$ and the transferred weight is large,
see Table I. The condition to activate spectral weight transfer between
the Hubbard subbands is finite hole doping in the LHB of the correlated 
$d$ band. Indeed, this hole doping nnd finite density within the $s$ 
orbitals induces the spectral transfer towards the LHB in the correlated 
band.

Altogether, the densities of states $D(\omega)$ give a metallic regime
for intermediate ($\alpha=0.5$) and strong ($\alpha=0.2$) screening of
strongly correlated $x^2-y^2$ states, see Fig.~\ref{fig:dos8}. A~large
gap between the Hubbard subbands opens when \mbox{$U_d=8$ eV;} this gap
is reduced to $\sim 0.5$ eV when $U_d=4.0$ eV. Nevertheless the system
has still an insulating gap which separates the Hubbard subbands. The 
electronic structure for the $x^2-y^2$ band is typical for a doped Mott 
insulator, with the weight of the UHB reduced by the kinetic processes 
in a doped system \cite{EskPRL,Esk94}. Indeed, the weight in the LHB 
above the Fermi energy increases by $\sim 2\delta$ where $\delta$ stands 
for the doping of the LHB, what would also be the weight transferred 
from the UHB to the LHB by finite doping. In this regime the $s$ band
is only weakly correlated and Hubbard subbands are poorly visible.

\subsection{Superconductivity in the two-band model}
\label{sec:sc}

\begin{figure}[t!]
	\includegraphics[width=\columnwidth]{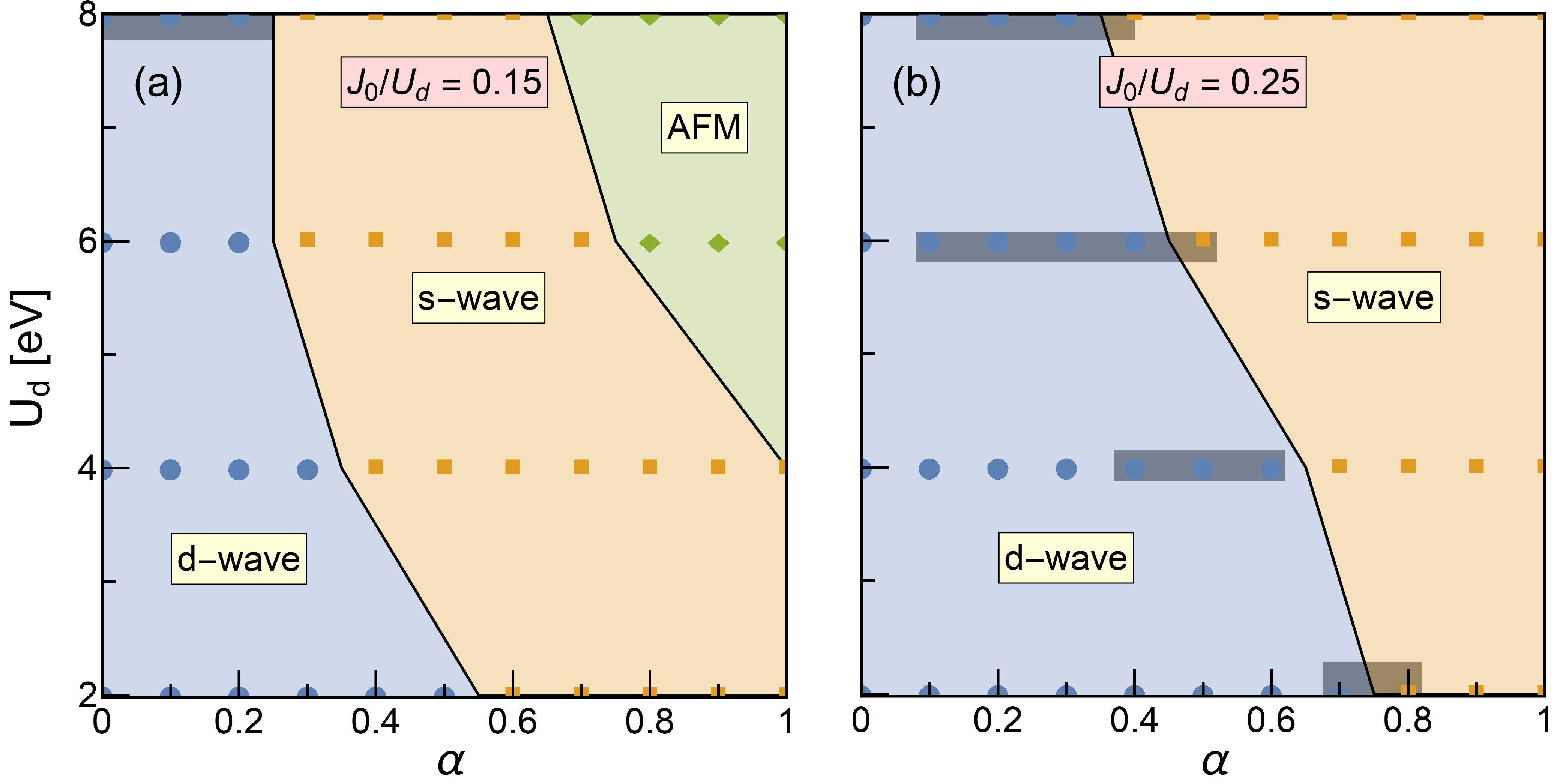}
	\caption{
Phase diagram in $(U_d,\alpha)$ plane for increasing
Hund's exchange, i.e., increasing ratio $J_0/U_d$:
(a) $J_0/U_d=0.15$ (as used above); (b) $J_0/U_d=0.25$. The pairing
has $s$- and $d$-wave symmetry, depending on the parameters.
Tendency towards triplet pairing is highlighted by gray boxes; 
the points mark the parameters at which the calculations were 
performed. 
}
	\label{fig:phd}
\end{figure}

Finally, we investigate the nature of the SC state. Therefore, we first 
compute the ground state of undoped two-band model on a finite square
cluster, i.e., for 8 electrons on 8-site cluster via exact (Lanczos) 
diagonalization. Next, the above cluster model is doped by 2 holes (by 
removing 2 electrons) and we look again at its ground state. After that 
pairing operators (for $s$- or $d$-wave) are applied on the undoped 
ground state and the overlap between the two states is obtained.
We consider SC states of both symmetries by changing the Coulomb 
interaction $U_d$ and the screening $\alpha$ in Fig.~\ref{fig:phd}.

For strong and unscreened Coulomb repulsion $U_d>4.0$ eV and weak Hund's 
exchange coupling, see Fig. \ref{fig:phd}(a), one finds AFM order in the 
undoped system with no sign of pairing. This regime resembles the state 
of cuprates: the $s$-like band is empty while the strongly correlated 
$x^2-y^2$ band is half-filled and Mott insulating, see 
Fig.~\ref{fig:mu}. This changes for weaker correlations (intermediate 
screening), where AFM order disappears and is replaced by $s$-wave 
pairing, see Fig.~\ref{fig:phd}(b).

The presence of $d$-wave pairing known from cuprates requires strong
screening. In the regime of strong screening doped holes enter the 
$x^2-y^2$ band and the model becomes similar to cuprate model. 
Altogether, Fig.~\ref{fig:phd} tells that stronger Hund's exchange 
promotes triplet pairing, reduces effective correlations, 
and suppresses AFM order.

In the phase diagram of Fig.~\ref{fig:phd}(a), AFM phase and $s$-wave
or $d$-wave pairings are accompanied by some indications of triplet
pairing. As expected, the latter is more pronounced at stronger Hund's
exchange coupling, see Fig.~\ref{fig:phd}(b). Energies obtained for the
doping with either one $\uparrow$ or one $\downarrow$ hole are here 
degenerate with the energies obtained with two $\uparrow$ holes, 
indicating that the doped hole-pair is a triplet. In order to check the 
stability of this result, we used again TBC. The degeneracy is then 
lifted and the $S^z=0$ state has lower energy, suggesting that triplet 
pairing might be a finite-size effect. Finally, we remark that 
interaction screening reduces the stability of AFM order in the 
correlated band and broadens the range of stability of $s$-wave and 
$d$-wave pairings, see Fig. \ref{fig:phd}. Moreover, the needed Hund's 
exchange to stabilize $d$-wave pairing in a broad regime is rather 
large ($J_0/U_d\gtrsim 0.25$).

\section{Summary and conclusions}
\label{sec:summa}

In summary, we have used exact diagonalization to investigate an
effective two-band model for infinite-layer nickelates, where the band
with a strong Ni($d_{x^2-y^2}$) character can be expected to be more
correlated than the one with a rather extended $s$-like wave function
of mostly rare-earth character. We focused here on the interactions in
both bands, especially their relative strength, which also tunes the
interorbital interactions between the two orbitals~\cite{Adh20}.
The latter give interband interactions and could generate
superconducting pairing.

We have established that both the very strongly correlated and the
strongly screened regimes support the mapping of the two-band model
onto a single Hubbard-like band. For (unrealistically) strong
interaction $U_d$, we find an antiferromagnetic Mott insulator without
tendencies to superconductivity. In the more realistic screened regime,
the $s$-like band takes up some of the charge carriers and the states
from both bands contribute at the Fermi energy. In this way the
correlated $x^2-y^2$ band~\cite{Kit20} is partly filled and the
spectral weight may be transferred to the unoccupied part of the lower
Hubbard band \cite{EskPRL}.

For intermediate screening the model is very rich and the $s$-band
hosts the doped holes forming $s$-wave pairs. We point out that this
situation broadly corresponds to a Kondo-lattice--like scenario, with
the caveat that the 'localized' $d_{x^2-y^2}$ spins can also move
\cite{Hep20,ZYZ20,Wan20p}. Hund's exchange coupling naturally yields
ferromagnetic interaction between itinerant $s$ carriers and
$d_{x^2-y^2}$ spins, but it is interesting to note that $s$-wave
pairing at stronger coupling was also obtained in a similar effective
model with AFM spin-spin coupling~\cite{Wan20p}. Altogether, this shows
that the model investigated here is very rich and predicts pairing of
different symmetry.

\acknowledgments

We thank Wojtek Brzezicki and Andres Greco for very insightful
discussions. T.~Plienbumrung acknowledges Development
and Promotion of Science and Technology Talents Project (DPST).
A.~M.~Ole\'s acknowledges Narodowe Centrum Nauki (NCN, Poland)
Project No. 2021/43/B/ST3/02166 and is grateful for support via
the Alexander von Humboldt Foundation Fellowskip
\mbox{(Humboldt-Forschungspreis).}

%

\end{document}